\documentclass[journal, draftclsnofoot, 11pt, a4paper, onecolumn]{IEEEtran}
\usepackage{graphicx}
\usepackage{amssymb}
\usepackage{amsmath}
\usepackage{calc}
\usepackage{array}
\usepackage{subfigure}
\usepackage{graphicx}
\usepackage{algorithm}
\usepackage{algorithmic}

\begin{document}
\title{\huge Linear Degrees of Freedom for $K$-user MISO Interference Channels with Blind Interference Alignment}

\author{Heecheol Yang,~\IEEEmembership{Student Member,~IEEE}, Wonjae Shin,~\IEEEmembership{Student Member,~IEEE},
and Jungwoo Lee,~\IEEEmembership{Senior Member,~IEEE} \vspace{-3mm}
\thanks{H. Yang, W. Shin, and J. Lee are with Wireless Signal Processing Lab., Dept. of Electrical and Computer 
Engineering, Seoul National Univ., Korea (e-mail:
hee2070@wspl.snu.ac.kr, \{wonjae.shin, junglee\}@snu.ac.kr).

This research was supported in part by Samsung Electronics Co., Ltd, Basic Science Research Program (NRF-2013R1A1A2008956) 
through the NRF funded by the MEST, Bio-Mimetic Robot Research Center funded by Defense Acquisition Program Administration (UD130070ID),
INMAC, and BK21-plus.
}
}

\markboth{IEEE Transactions on Wireless Communications, Draft}{YANG, et al.:
Linear Degrees of Freedom for $K$-user MISO Interference Channels with Blind Interference Alignment}

\maketitle

\begin{abstract}
In this paper, we characterize the degrees of freedom (DoF) for $K$-user $M \times 1$ multiple-input single-output interference channels with reconfigurable antennas which have multiple preset modes at the receivers, assuming linear coding strategies in the absence of channel state information at the transmitters, i.e., blind interference alignment.
Our linear DoF converse builds on the lemma that if a set of transmit symbols is aligned at their common unintended receivers, those symbols must have independent signal subspace  at their corresponding receivers.
This lemma arises from the inherent feature that channel state's changing patterns of the links towards the same receiver are always identical, assuming that the coherence time of the channel is long enough.
We derive an upper bound for the linear sum DoF, and 
propose an achievable scheme that exactly achieves the linear sum DoF upper-bound when both of the $\frac{n^{*}}{M}=R_{1}$ and $\frac{MK}{n^{*}}=R_{2}$ are integers.
For the other cases, where either $R_{1}$ or $R_{2}$ is not an integer, we only give some guidelines how the interfering signals are aligned at the receivers to achieve the upper-bound.
As an extension, we also show the linear sum DoF upper-bound for downlink/uplink cellular networks.
\end{abstract}


\section{Introduction}
\label{sec_intro}
\PARstart{I}{nterference} alignment (IA) has attracted much attention due to its novel approach 
to interference-limited networks \cite{IA}-\cite{XIA}. This scheme aligns interfering signals into small subspaces
to leave room for the desired signal dimensions. For $K$-user interference channels (IC), 
each user achieves 1/2 degrees of freedom (DoF) by the IA approach. 
In this scheme, global channel state information at transmitter (CSIT)
is necessary to align interfering signals at unintended receivers. 
However, global CSIT is hard to achieve or even cannot be achieved 
in pracical systems. Even when global CSIT is available, 
transmitters may obtain imperfect channel knowledge due to quantization error and feedback delay.

In the absence of CSIT, interference-limited networks cannot achieve the sum DoF more than 1 for a general channel state condition \cite{Huang}-\cite{Vaze}.
However, it was introduced that the IA technique increases the sum DoF without CSIT for a specific channel's coherence time/bandwidth condition \cite{blind}-\cite{blind2}.
With such a channel condition, a total of $K/2$ DoF is achievable for $K$-user IC, which meets the 
outer bound of the perfect CSIT scenario. This IA technique is referred to as \emph{blind interference alignment (BIA)}.
To make the specific channel condition more feasible, BIA through 
reconfigurable antenna switching was proposed \cite{Gou}. 
The DoF characterization with reconfigurable antennas has been studied for various network scenarios with the assumption that
the number of preset modes at a receiver ($N$) is not greater than the number of transmit antennas ($M$), i.e., $M \geq N$ \cite{Gou}-\cite{IC2}.
According to \cite{Gou}, $K$-user $M \times 1$ multiple-input 
single-output (MISO) broadcast channels (BC) achieve a total of $\frac{MK}{M+K-1}$ DoF with $M$ preset modes of each reconfigurable antenna (i.e., $N_{i}=M, \forall i$, where receiver $i$ has $N_{i}$ preset modes). By the result of \cite{LDOF_MIMO}, it was demonstrated that the sum DoF for MISO BC 
cannot be increased even if a reconfigurable antenna has more than $M$ preset modes.
Meanwhile, for the IC scenario, a new achievable scheme for the $K$-user MISO IC is proposed in \cite{IC}-\cite{IC2} for the asymmetric antenna configuration,
with the condition that $N_{i} = M_{i}, \forall i$, where transmitter $i$ (receiver $i$) has $M_{i}$ ($N_i$) antennas (preset modes), respectively.
When transmitters are equipped with different number of transmit antennas, the number of transmit symbols is dependent on the number of transmit antennas.
If $M_{i}=M$ for all $i$, the
$K$-user $M \times 1$ MISO IC also achieves a total of $\frac{MK}{M+K-1}$ DoF as the $K$-user $M \times 1$ MISO BC.

The main contribution of this paper is to fully characterize the DoF for the $K$-user $M \times 1$ MISO IC 
with reconfigurable antennas at the receivers by considering only linear coding strategies without channel knowledge at transmitters, which includes the case where the number of preset modes is greater than the number of transmit antennas, i.e. $N > M$. This is the most
general result in the literature. It was recently reported that a total of $6/5$ DoF is achievable for the 3-user single-input single-output (SISO) IC with $2$ preset modes \cite{IC3}. This result shows that extra preset modes at the receivers in comparison with the 
number of transmit antennas can be exploited to increase the sum DoF for the IC scenario, while extra preset modes cannot have a beneficial effect on the sum DoF for the BC scenario \cite{LDOF_MIMO}.
Subsequently, it was simply extended in \cite{IC4} that the $K$-user SISO IC with 2 receiver preset modes  achieves a total of $\frac{2K}{K+2}$ DoF. In this paper, we derive a new linear sum DoF upper-bound for the $K$-user $M \times 1$ MISO IC with $N$ 
preset modes, including the $N > M$ scenario. 

In our approach to derive the linear sum DoF upper-bound, we focus on a lemma which claims that if a set of transmit symbols is aligned at their common unintended receivers, those symbols must have independent signal subspace at their intended receivers, which is induced from the characteristic of the channel matrix with reconfigurable antenna switching.
Specifically, we exploit the notable feature that channel states of the links towards the same receiver
must have the same changing pattern since the changing pattern of the channel state is determined solely by
the receiver's preset mode pattern, under the assumption that the coherence time of the channel is long enough.
Remarkably, we show that the linear sum DoF upper-bound is achievable when $\frac{n^{*}}{M}=R_{1}$ and $\frac{MK}{n^{*}}=R_{2}$ are integers, where $n^{*}$ represents the optimal number of preset modes among $N$ preset modes at each receiver, under the linear scheme obtained by modifying the existing achievable scheme for the IC scenario in \cite{IC2}. If these conditions are not satisfied, i.e., either $R_{1}$ or $R_{2}$ is not an integer, we give some guidelines for an achievable scheme, which shows how to align interfering signals by introducing the concept of an alignment set.

We also extend our approach to downlink/uplink cellular networks.
There has been previous work on the BIA scheme with reconfigurable antennas for cellular networks:
the cluster-based frequency reuse system \cite{cluster1}-\cite{cluster2}, data sharing for cell-edge users \cite{heath}-\cite{cespedes1}, and the sum DoF characterizations \cite{cespedes2}-\cite{wei}.
In this paper, we derive the linear sum DoF upper-bound for fully connected downlink cellular networks with reconfigurable antennas at the users, by applying the same lemma for the IC scenario.
We also show this upper-bound implies that extra preset modes at the users compared with base station's antennas can be useful to increase the linear sum DoF without data sharing between base stations.
In addition, our result is applied to uplink scenario with reconfigurable antennas at the base station by considering the effect of transmitter cooperation for the IC scenario with no CSIT.

The rest of this paper is organized as follows.
In Section \ref{sec_model}, we introduce the system model and the main results of this paper. 
In Section \ref{sec_DoF}, the linear sum DoF converse for the $K$-user MISO IC is proved.
We propose a new achievable scheme that exactly achieves the linear sum DoF upper-bound in Section \ref{sec_ach}. 
In Section \ref{sec_cell}, we also discuss the linear sum DoF for downlink/uplink cellular networks.
We conclude this paper in Section \ref{sec_con}.

\emph{Notation}:
For a vector $\mathbf{a}$, 
$\textrm{diag}(\mathbf{a})$ represents a diagonal matrix whose diagonal entries are elements of $\mathbf{a}$.
For matrices $\mathbf{A}$ and $\mathbf{B}$, $\mathbf{A}^{T}$ means the transpose of $\mathbf{A}$, 
$\textrm{span}(\mathbf{A})$ denotes the space spanned by the column vectors of $\mathbf{A}$, 
$\textrm{dim}(\mathbf{A})$ is the dimension of $\textrm{span}(\mathbf{A})$, and
$\textrm{dim}(\mathbf{A} \cap \mathbf{B})$ represents the dimension of the intersection of $\textrm{span}(\mathbf{A})$ and $\textrm{span}(\mathbf{B})$.
For vector spaces $\mathcal{A}$ and $\mathcal{B}$, $\textrm{Proj}_{\mathcal{A}^{c}}\mathcal{B}$ denotes the vector space induced by projecting $\mathcal{B}$ onto the orthogonal complement of $\mathcal{A}$.
For $r \in \mathbb{R}$, $\lfloor r \rfloor$ is the largest integer not greater than $r$,
and $\lceil r \rceil$ is the smallest integer not less than $r$.
For $a,b \in \mathbb{N}$, $[a:b]$ denotes $\{a, a+1, \ldots, b\}$. 
For sets $\mathcal{A}$ and $\mathcal{B}$, $\mathcal{A}\setminus\mathcal{B}$ is the relative complement of $\mathcal{B}$ in 
$\mathcal{A}$. 
The notation $\binom{n}{k}$ is a $k$-combination of a set $\mathcal{S}$, which has $n$ elements.
\section{System Model \& Main Results}
\label{sec_model}
\begin{figure}[t]
    \centerline{\includegraphics[width=8.0cm]{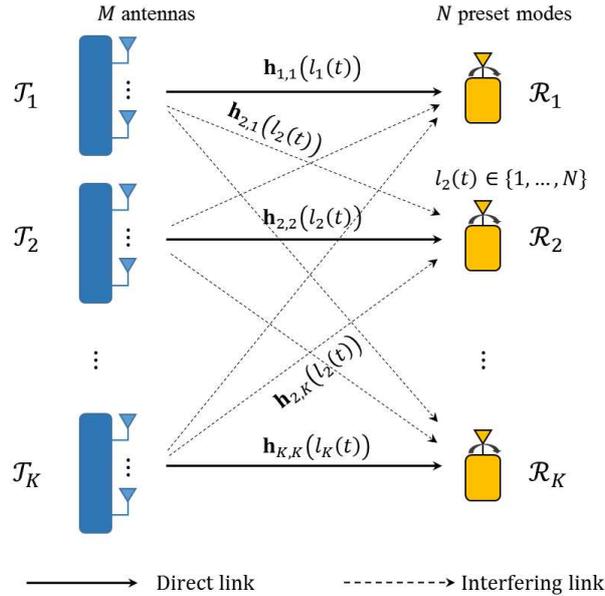}}
    \caption{System model for $K$-user $M \times 1$ interference channels with $N$ preset modes at the receivers.}
    \label{SysModel}
    \vspace{-3mm}
\end{figure}

Consider the $K$-user $M \times 1$ MISO IC with a reconfigurable antenna switching.
The system has $K$ transmitters, each of which has $M$ transmit antennas, and $K$ receivers which have
a single reconfigurable antenna with $N$ preset modes each.
We refer to it as $(M,N,K)$-IC from now on, which is described in Fig. \ref{SysModel}.
We notate $\mathcal{T}_{\mathcal{S}}$ ($\mathcal{R}_{\mathcal{S}}$) to represent a set of transmitters (receivers) for $\mathcal{S} \subset [1:K]$, 
e.g., $\mathcal{T}_{[1:3]}=\{\textrm{Transmitter } 1, \textrm{Transmitter } 2, \textrm{Transmitter } 3 \}$.
It is simply notated as $\mathcal{T}$ ($\mathcal{R}$) and $\mathcal{T}_{a}$ ($\mathcal{R}_{a}$) to denote 
$\mathcal{T}_{[1:K]}$ ($\mathcal{R}_{[1:K]}$) and Transmitter $a$ (Receiver $a$), respectively.
At $\mathcal{R}_{j}$, the transmit signals from $\mathcal{T}_{j}$ are desired signals,
while those from other transmitters (i.e., $\mathcal{T} \setminus \mathcal{T}_{j}$) are interfering signals.
We denote the $a^{th}$ antenna of $\mathcal{T}_{i}$ by $i(a)$ for $i \in [1:K]$ and $a \in [1:M]$, and $\mathcal{B}$ denotes a set of all transmit antennas in the network as $\mathcal{B}=\{1(1),\ldots, K(M)\}$. 
The preset mode of $\mathcal{R}_{j}$ at time $t$ is denoted by $l_{j}(t)$.
We assume that channel coefficients remain constant during the symbol extension period, i.e. the coherence time of the channel is long enough.
In this scenario, since the channel state varies according to the receiver's preset mode, we represent the channel vector
from $\mathcal{T}_{i}$ to $\mathcal{R}_{j}$ at time $t$ as $\mathbf{h}_{j,i}(l_{j}(t)) =  [h_{j,i(1)}(l_{j}(t)) \ldots h_{j,i(M)}(l_{j}(t))] \in \mathbb{C}^{1 \times M}$, 
each entry of which is assumed to be independent and identically distributed (i.i.d.).
Then, the channel matrix from the transmit antenna $i(a)$ to $\mathcal{R}_{j}$ over $m$ channel uses is denoted by
$\mathbf{H}^{m}_{j,i(a)}=\textrm{diag}([h_{j,i(a)}(l_{j}(1)) \ldots h_{j,i(a)}(l_{j}(m))])$.
Meanwhile, $\mathcal{T}_{i}$ sends $\sum\nolimits_{a=1}^{M} d_{i(a)}$ symbols, and the 
transmit symbols of the transmit antenna $i(a)$ ($a \in [1:M]$) over $m$ channel uses is
$\mathbf{x}^{m}_{i(a)}=\sum\nolimits_{d=1}^{d_{i(a)}}s_{i(a),d}\mathbf{v}^{m}_{i(a),d}$
where $s_{i(a),d}$ is the $d^{th}$ data symbol from transmit antenna $i(a)$ and $\mathbf{v}^{m}_{i(a),d} \in \mathbb{C}^{m \times 1}$ is the transmit beamforming vector for $s_{i(a),d}$ which is comprised of 0 and 1.
We also denote the beamforming matrix of transmit antenna $i(a)$ over $m$ channel uses as
$\mathbf{V}^{m}_{i(a)}=[\mathbf{v}^{m}_{i(a),1} \ldots \mathbf{v}^{m}_{i(a),d_{i(a)}}]$.
When $M=1$, we simplify the notations of $\mathbf{v}^{m}_{i(a),d}$ and $s_{i(a),d}$ as $\mathbf{v}^{m}_{i,d}$ and $s_{i,d}$, respectively.
The received signal at $\mathcal{R}_{j}$ over $m$ channel uses is
\begin{eqnarray}
\mathbf{y}^{m}_{j}=\underbrace{\sum\limits_{a=1}^{M}\mathbf{H}^{m}_{j,j(a)}\mathbf{x}^{m}_{j(a)}}_{\textrm{desired signals}}
+ \underbrace{\sum\limits_{i=1, i\neq j}^{K}\sum\limits_{a=1}^{M}\mathbf{H}^{m}_{j,i(a)}\mathbf{x}^{m}_{i(a)}}_
{\textrm{interfering signals}}+\mathbf{z}^{m}_{j},
\end{eqnarray}
where $\mathbf{z}^{m}_{j} \in \mathbb{C}^{m \times 1}$ is the additive white Gaussian noise over $m$ channel uses, each entry of which is distributed as $\mathcal{CN}(0,1)$.
The transmitters are subject to the average transmit power constraint $P$.

According to the BIA concept using reconfigurable antennas proposed in \cite{Gou}, interfering signals are aligned
through a predetermined order of antenna switching, which is called preset mode pattern.
We define the preset mode pattern of $\mathcal{R}_{j}$ during $m$ channel uses as $\mathbf{L}^{m}_{j}=[l_{j}(1) \ldots l_{j}(m)]$.
Since we assume that channel coefficients remain constant during a symbol extension period,
the channel state varies depending on the receiver's preset mode pattern.
Therefore, the channel states of links towards the same receiver have the same changing pattern.
In addition, the system assumes that CSIT is not available, and receivers have perfect channel knowledge.
Lastly, the linear DoF (LDoF) of $K$-tuple $(d_{1},\ldots,d_{K})$ is achievable if there exists a set of beamforming vectors and preset mode patterns for $j \in [1:K]$ almost surely, satisfying
\begin{eqnarray}
\label{D_tuple}
\begin{array}{c}
\hspace{-3mm} \textrm{dim}\left(\textrm{Proj}_{\mathcal{I}^{c}_{j}}\textrm{span}([\mathbf{H}^{m}_{j,j(1)}\mathbf{V}^{m}_{j(1)} \cdots \mathbf{H}^{m}_{j,j(M)}\mathbf{V}^{m}_{j(M)}])\right)=d_{j}(m), \\ d_{j}=\underset{m\to\infty}{\textrm{lim}}\frac{d_{j}(m)}{m},
\end{array} \hspace{-5mm}
\end{eqnarray}
where $\mathcal{I}_{j}$ is the interference signal subspace at $\mathcal{R}_{j}$ as 
\begin{eqnarray}
\mathcal{I}_{j}&=&\textrm{span}\Big([\mathbf{H}^{m}_{j,1(1)}\mathbf{V}^{m}_{1(1)} \cdots \mathbf{H}^{m}_{j,j-1(M)}\mathbf{V}^{m}_{j-1(M)} \\ \nonumber
&& \hspace{10mm} \mathbf{H}^{m}_{j,j+1(1)}\mathbf{V}^{m}_{j+1(1)} \cdots \mathbf{H}^{m}_{j,K(M)}\mathbf{V}^{m}_{K(M)}]\Big).
\end{eqnarray}
The LDoF region $\mathcal{D}$ is the closure of the set of all achievable LDoF tuples satisfying (\ref{D_tuple}), and 
the linear sum DoF is given by
\begin{eqnarray}
\textrm{LDoF}_{\textrm{sum}}=\underset{(d_{1},\ldots,d_{K}) \in \mathcal{D}}{\mathrm{max}}\sum\limits_{j=1}^{K}d_{j}.
\end{eqnarray}

The main results on linear sum DoF upper-bound and achievability are as follows.

\textbf{Theorem 1} (DoF converse) :
For the $K$-user $M\times1$ MISO IC with $N$ preset modes at each receiver, 
the linear sum DoF is upper bounded by
\begin{eqnarray}
\label{eq:DoF_n}
\textrm{LDoF}_{\textrm{sum}} \leq \frac{n^{*}K}{K+\left\lceil\frac{n^{*}}{M}\right\rceil(n^{*}-1)},
\end{eqnarray}
where $N=M\Gamma+\alpha$ ($0 \leq \alpha < M$), and $n^{*}$ represents the optimal number of preset modes among $N$ preset modes as
\begin{eqnarray} \label{eq:n}
n^{*}=\left\{\begin{array}{cc} 
M\Gamma, & N < M \left\lceil\sqrt{\frac{K}{M}}\right\rceil, \alpha \leq \frac{N(M\Gamma-1)}{K-\Gamma-1} \\
N, & N < M \left\lceil\sqrt{\frac{K}{M}}\right\rceil, \alpha > \frac{N(M\Gamma-1)}{K-\Gamma-1} \\
M \Gamma_{\textrm{opt}}, & N \geq M \left\lceil\sqrt{\frac{K}{M}}\right\rceil \end{array} \right.
\end{eqnarray}
where $\Gamma_{\textrm{opt}}=\underset{\gamma = \left\lfloor\sqrt{\frac{K}{M}}\right\rfloor, \left\lceil\sqrt{\frac{K}{M}}\right\rceil}
{\mathrm{argmin}} M\gamma+\frac{K}{\gamma}$.

\textbf{Theorem 2} (DoF achievability):
For the $K$-user $M \times 1$ MISO IC with $N$ preset modes at each receiver,
if $R_{1}$ and $R_{2}$ are integers where $R_{1}=\frac{n^{*}}{M}$, $R_{2}=\frac{MK}{n^{*}}$, and
$n^{*}$ is determined by (\ref{eq:n}), 
the linear sum DoF upper-bound (\ref{eq:DoF_n}) is achievable when beamforming vectors and 
preset mode patterns are constructed as $R_{2}$-user $n^{*} \times 1$ MISO IC according to \cite{IC2}.

\textbf{Corollary 1}:
For $K$-user $M \times 1$ MISO IC with $N$ preset modes at each receiver,
the linear sum DoF is
\begin{eqnarray}
\textrm{LDoF}_{\textrm{sum}} = \frac{n^{*}K}{K+\left\lceil\frac{n^{*}}{M}\right\rceil(n^{*}-1)},
\end{eqnarray}
where $n^{*}$ is determined by (\ref{eq:n}), $n^{*}$ is a multiple of $M$, and $MK$ is divisible by $n^{*}$.
\begin{proof}
It is simply induced from Theorem 1 and 2.
\end{proof}

We prove Theorem 1 and 2 in Section \ref{sec_DoF} and \ref{sec_ach}, respectively.
\section{DoF Converse}
\label{sec_DoF}

In this section, we derive the upper-bound of the linear sum DoF for the $K$-user $M \times 1$ MISO IC with reconfigurable antennas equipped with $N$ preset modes at the receivers.
To begin with, we introduce key lemmas to prove the DoF converse, which is followed by the proof of Theorem 1.
In addition, we discuss the linear sum DoF upper-bound tendency as $N$ increases, and compare this result with finite state compound wireless network scenarios.

\subsection{Key Lemmas}
\label{sec_lemma}
\begin{figure*}[t]
    \centerline{\includegraphics[width=15.0cm]{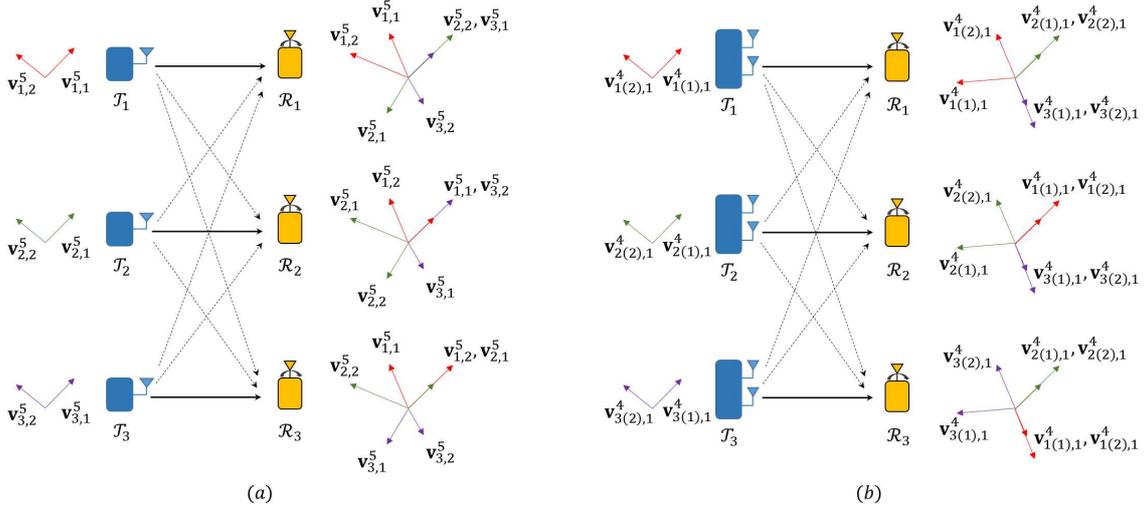}}
    \caption{(a) Description for the 3-user SISO IC with 2 preset modes, $(1,2,3)$-IC. The aligned beamforming vectors at the receivers represent the alignment set which is aligned in the 1-dimensional signal subspace at that receiver.
    (b) Description for $(2,2,3)$-IC.}
    \label{3SISO}
    \vspace{-3mm}
\end{figure*}

Beforehand, we introduce a lemma in \cite{IC3} that shows an interesting feature of channel matrices with reconfigurable antenna switching at receivers, which comes from the property of diagonal matrices.

\textbf{Lemma 1} (Lemma 2 in \cite{IC3}):
If $\mathbf{H}_{1}\mathbf{v}_{1}\in$ span($\mathbf{H}_{2}\mathbf{V}_{2}$), where $\mathbf{H}_{1}$ and $\mathbf{H}_{2}$ are two $m \times m$ full-rank diagonal matrices with the same pattern, implying that their diagonal entries have the same changing pattern, $\mathbf{v}_{1}$ is an $m \times 1$ column vector, $\mathbf{V}_{2}$ is an $m \times d$ thin matrix, i.e., $d<m$, 
and the entries of $\mathbf{v}_{1}$ and $\mathbf{V}_{2}$ are generated independently of the values of the diagonal entries of 
$\mathbf{H}_{1}$ and $\mathbf{H}_{2}$, then $\mathbf{v}_{1} \in$ span($\mathbf{V}_{2}$).
\begin{proof}
We refer to Lemma 2 in \cite{IC3}.
\end{proof}

The underlying assumptions in this lemma for the vector and matrices (i.e., $\mathbf{H}_{1}, \mathbf{H}_{2}, \mathbf{v}_{1}, \mathbf{V}_{2}$) match up well with our system model. Since channel state varies according to the receiver's preset mode, it can be seen that $\mathbf{H}_{1}$ and $\mathbf{H}_{2}$ denote the channel matrices towards the same receiver from different transmitters since they have the same changing pattern in their diagonal entries. In addition, since $\mathbf{v}_{1}$ and $\mathbf{V}_{2}$ are generated independently of the channel values, they can represent the beamforming vector and matrix in our system model, respectively. Even when $\mathbf{v}_{1}$ and $\mathbf{V}_{2}$ are relevant to the changing pattern of  $\mathbf{H}_{1}$ and $\mathbf{H}_{2}$, not the channel realizations, Lemma 1 still holds. For example, if the transmit symbol from $\mathcal{T}_{i}$, $\mathbf{v}^{m}_{i(a),1}$ for $a \in [1:M]$, is aligned at $\mathcal{R}_{k}$ in the space spanned by the interfering signal from $\mathcal{T}_{j}$ with reconfigurable antenna switching, then it can be denoted as $\mathbf{H}^{m}_{k,i(a)}\mathbf{v}^{m}_{i(a),1}\in$ span($\mathbf{H}^{m}_{k,j(b)}\mathbf{V}^{m}_{j(b)}$) for $b \in [1:M]$. In this case, by Lemma 1, $\mathbf{v}^{m}_{i(a),1}$ should be included in the space spanned by $\mathbf{V}^{m}_{j(b)}$, i.e., $\mathbf{v}^{m}_{i(a),1} \in$ span($\mathbf{V}^{m}_{j(b)}$).

We extend Lemma 1 to show that a set of transmit symbol vectors should be aligned only at their common unintended receivers with reconfigurable antenna switching at each receiver in the absence of CSIT. 

\textbf{Lemma 2}:
When the transmit signals from $\mathcal{T}_{\mathcal{S}} \subset \mathcal{T}$ are aligned at $\mathcal{R} \setminus \mathcal{R}_{\mathcal{S}}$ by reconfigurable antenna switching at the receiver without any knowledge of channel realizations at the transmitter, 
any of those signals should not be aligned in the interference signal subspace at all $\mathcal{R}_{j} \subset \mathcal{R}_{\mathcal{S}}$ spanned by the other interfering signal from 
$\mathcal{T} \setminus \mathcal{T}_{j}$.
\begin{proof}
Without loss of generality, suppose that the transmit signals from $\mathcal{T}_{[1:n]}$ ($1\leq n<K$) are aligned at $\mathcal{R} \setminus \mathcal{R}_{[1:n]}$.
For instance, when $\mathbf{v}^{m}_{1(1),1}$ is aligned at all $\mathcal{R}_{k} \subset \mathcal{R} \setminus \mathcal{R}_{[1:n]}$ with other symbols from $\mathcal{T}_{[1:n]}$,
\begin{eqnarray}
\textrm{at $\mathcal{R}_{k}$: } \mathbf{H}^{m}_{k,1(1)}\mathbf{v}^{m}_{1(1),1} \in \textrm{span}(\mathbf{H}^{m}_{k,j(a)}\mathbf{V}^{m}_{j(a)}),
\end{eqnarray}
for $j \in [1:n]$ and $a \in [1:M]$.
Recall that transmitters have no knowledge about channel realizations, and $\mathbf{H}^{m}_{k,1(1)}$ and $\mathbf{H}^{m}_{k,j(a)}$ have the identical changing pattern of their diagonal entries. By Lemma 1,
\begin{eqnarray}
\mathbf{v}^{m}_{1(1),1} \in \textrm{span}(\mathbf{V}^{m}_{j(a)}),
\label{eq:v_1}
\end{eqnarray}
At the same time, if $\mathbf{v}^{m}_{1(1),1}$ is aligned in the interference signal subspace at $\mathcal{R}_{j} \subset \mathcal{R}_{[1:n]}$ spanned by the transmit signals from $\mathcal{T}_{i} \subset \mathcal{T} \setminus \mathcal{T}_{j}$,
\begin{eqnarray}
\textrm{at $\mathcal{R}_{j}$: } \mathbf{H}^{m}_{j,1(1)}\mathbf{v}^{m}_{1(1),1} \in \textrm{span}(\mathbf{H}^{m}_{j,i(b)}\mathbf{V}^{m}_{i(b)}),
\end{eqnarray}
for $b \in [1:M]$.
This expression can be written as
\begin{eqnarray}
\textrm{at $\mathcal{R}_{j}$: } \mathbf{H}^{m}_{j,1(1)}\mathbf{v}^{m}_{1(1),1} \in 
\textrm{span}(\mathbf{H}^{m}_{j,j(a)}(\mathbf{H}^{m}_{j,j(a)})^{-1}\mathbf{H}^{m}_{j,i(b)}\mathbf{V}^{m}_{i(b)}). \hspace{-2mm}
\end{eqnarray}
Since $\mathbf{H}^{m}_{j,1(1)}$ and $\mathbf{H}^{m}_{j,j(a)}$ have the same diagonal changing pattern,
\begin{eqnarray}
\mathbf{v}^{m}_{1(1),1} \in \textrm{span}((\mathbf{H}^{m}_{j,j(a)})^{-1}\mathbf{H}^{m}_{j,i(b)}\mathbf{V}^{m}_{i(b)}),
\label{eq:v_2}
\end{eqnarray}
by Lemma 1. From (\ref{eq:v_1}) and (\ref{eq:v_2}), since $\mathbf{v}_{1(1),1}$ is included in both of the two vector spaces span$(\mathbf{V}^{m}_{j(a)})$ and span$((\mathbf{H}^{m}_{j,j(a)})^{-1}\mathbf{H}^{m}_{j,i(b)}\mathbf{V}^{m}_{i(b)})$, they have a non-zero dimensional intersection of the vector space. Thus,
\begin{eqnarray}
\textrm{dim}(\mathbf{V}^{m}_{j(a)} \cap (\mathbf{H}^{m}_{j,j(a)})^{-1}\mathbf{H}^{m}_{j,i(b)}\mathbf{V}^{m}_{i(b)}) > 0,
\end{eqnarray}
and it leads to
\begin{eqnarray}
\textrm{dim}(\mathbf{H}^{m}_{j,j(a)}\mathbf{V}^{m}_{j(a)} \cap \mathbf{H}^{m}_{j,i(b)}\mathbf{V}^{m}_{i(b)}) > 0.
\end{eqnarray}
It means that $\mathbf{H}^{m}_{j,j(a)}\mathbf{V}^{m}_{j(a)}$ and $\mathbf{H}^{m}_{j,i(b)}\mathbf{V}^{m}_{i(b)}$ have the intersection of the signal subspace at $\mathcal{R}_{j}$.
For $\mathcal{R}_{j}$, $\mathbf{H}^{m}_{j,j(a)}\mathbf{V}^{m}_{j(a)}$ is a desired signal from $\mathcal{T}_{j}$, while 
$\mathbf{H}^{m}_{j,i(b)}\mathbf{V}^{m}_{i(b)}$ is an interference signal.
Since $\mathbf{H}^{m}_{j,j(a)}\mathbf{V}^{m}_{j(a)}$ is contaminated by the interference signal from $\mathcal{T}_{i}$,
it can be said that $\mathbf{v}^{m}_{1(1),1}$ should not be aligned in the interference signal subspace at all $\mathcal{R}_{j}\subset \mathcal{R}_{[1:n]}$ spanned by the transmit signal from 
$\mathcal{T} \setminus \mathcal{T}_{j}$ when it is aligned at $\mathcal{R} \setminus \mathcal{R}_{[1:n]}$ with the other transmit symbols from $\mathcal{T}_{[1:n]}$ to guarantee the interference-free signal subspace.
\end{proof}

Lemma 2 implies that if transmit symbols from $\mathcal{T}_{\mathcal{S}}$ are aligned in the interference signal subspaces at their common unintended receivers, i.e., 
$\mathcal{R} \setminus \mathcal{R}_{\mathcal{S}}$,
each of those symbols has independent subspace at $\mathcal{R}_{\mathcal{S}}$,
although each symbol is only desired by one intended receiver.
Motivated by Lemma 2, we define the notion of an \emph{alignment set}.

\textbf{Definition}: If a set of transmit symbols from $\mathcal{T}_{S}$ are aligned in the 1-dimensional signal subspace at their common unintended receivers, $\mathcal{R} \setminus \mathcal{R}_{\mathcal{S}}$, while occupying independent signal subspaces at each of their corresponding receivers, $\mathcal{R}_{\mathcal{S}}$, they are called \emph{alignment set $\mathcal{A}_{S}$}.  

Using this terminology, we can equivalently reinterpret the existing linear sum DoF results for SISO IC and MISO IC scenarios introduced in \cite{IC2} and \cite{IC3}, respectively.

\emph{Example 1}:
Consider 3-user SISO IC with two preset modes, i.e., $(1,2,3)$-IC, which achieves a total of 6/5 LDoF \cite{IC3} in Fig \ref{3SISO}-(a).
Each user sends two symbols over 5-symbol extension, 
thereby $\mathbf{v}^{5}_{1,1}$ and $\mathbf{v}^{5}_{3,2}$ are aligned in an 1-dimensional subspace at $\mathcal{R}_{2}$,
$\mathbf{v}^{5}_{1,2}$ and $\mathbf{v}^{5}_{2,1}$ are aligned in an 1-dimensional subspace at $\mathcal{R}_{3}$, 
and $\mathbf{v}^{5}_{2,2}$ and $\mathbf{v}^{5}_{3,1}$ are also aligned in an 1-dimensional subspace at $\mathcal{R}_{1}$.
It can be said that the alignment sets are constructed as
\begin{eqnarray}
\begin{array}{c}
\mathcal{A}_{\{1,3\}}=\{\mathbf{v}^{5}_{1,1}, \mathbf{v}^{5}_{3,2}\}, \\
\mathcal{A}_{\{1,2\}}=\{\mathbf{v}^{5}_{1,2}, \mathbf{v}^{5}_{2,1}\}, \\
\mathcal{A}_{\{2,3\}}=\{\mathbf{v}^{5}_{2,2}, \mathbf{v}^{5}_{3,1}\}.
\end{array}
\end{eqnarray}
At $\mathcal{R}_{1}$, although $\mathbf{v}^{5}_{2,1}$ and $\mathbf{v}^{5}_{3,2}$ are interfering symbol vectors,
they occupy an independent 1-dimensional subspace each, since each of them is aligned with the different transmit symbol vectors from $\mathcal{T}_{1}$ at the unintended receivers.
In a similar way, because each receiver sets aside 2-dimensional signal subspace for the desired signals among a total of 5-dimensional signal space,
it is observed that a total of 6/5 LDoF is achievable for 3 users.

\emph{Example 2}:
Consider another MISO IC scenario, $(2,2,3)$-IC, which achieves a total of 3/2 LDoF \cite{IC2} in Fig \ref{3SISO}-(b).
In this scenario, the transmit symbol vectors from each transmitter's two antennas are aligned in 1-dimensional signal subspace at other receivers. During 4-symbol extension, the transmit symbol vectors $\mathbf{v}^{4}_{1(1),1}$ and $\mathbf{v}^{4}_{1(2),1}$ can be aligned at $\mathcal{R} \setminus \mathcal{R}_{1}$ since they are desired to only $\mathcal{R}_{1}$. 
By the same manner, $\mathbf{v}^{4}_{2(1),1}$ and $\mathbf{v}^{4}_{2(2),1}$ are aligned at $\mathcal{R} \setminus \mathcal{R}_{2}$,
and $\mathbf{v}^{4}_{3(1),1}$ and $\mathbf{v}^{4}_{3(2),1}$ are aligned at $\mathcal{R} \setminus \mathcal{R}_{3}$.
In this case, each of the alignment sets include the transmit symbols intended to a single receiver as
\begin{eqnarray}
\begin{array}{c}
\mathcal{A}_{\{1\}}=\{\mathbf{v}^{4}_{1(1),1}, \mathbf{v}^{4}_{1(2),1}\}, \\
\mathcal{A}_{\{2\}}=\{\mathbf{v}^{4}_{2(1),1}, \mathbf{v}^{4}_{2(2),1}\}, \\
\mathcal{A}_{\{3\}}=\{\mathbf{v}^{4}_{3(1),1}, \mathbf{v}^{4}_{3(2),1}\}.
\end{array}
\end{eqnarray}
With these alignment sets, each user gets 2 desired symbols during 4-symbol extension.
Consequently, each user achieves 2/4 LDoF, so that a total of 3/2 LDoF can be achieved. Compared to Example 1, each user achieves a greater DoF by using two transmit antennas.

With regard to the alignment set, we claim an important lemma to derive the upper-bound.

\textbf{Lemma 3}: The alignment sets have the same cardinality to maximize the sum DoF upper-bound.
\begin{proof}
We defer the proof into Appendix.
\end{proof}

According to Lemma 3, we only consider the symmetric case for the cardinality of the alignment sets. Thus, all alignment sets are assumed to have $n$ transmit symbols. 
In addition, we introduce a supplementary lemma that could be used to induce the linear DoF from the dimension of signal subspaces. 

\textbf{Lemma 4} (Lemma 3 in \cite{Lash}): For two matrices $\mathbf{A}$ and $\mathbf{B}$ with the same row size,
\begin{eqnarray}
\textrm{dim}\left(\textrm{Proj}_{\mathcal{A}^{c}} \mathcal{B}\right) = \textrm{dim} ([\mathbf{A} \hspace{1mm} \mathbf{B}]) - \textrm{dim}(\mathbf{A}),
\end{eqnarray}
where $\mathcal{A}$ and $\mathcal{B}$ denotes $\textrm{span}(\mathbf{A})$ and $\textrm{span}(\mathbf{B})$, respectively.
It can be derived by basic linear algebra, thus its proof is omitted.
We now extend the approach inspired by Lemma 2 to prove the linear sum DoF converse for the general $(M,N,K)$-IC scenario.


\subsection{Proof of Theorem 1}
\label{sec_pf1}
From the definition of $d_{j}(m)$ and Lemma 4,
\begin{eqnarray}
\label{eq:lemma3}
d_{j}(m) \hspace{-3mm} &=& \hspace{-3mm} \textrm{dim}\left(\textrm{Proj}_{\mathcal{I}^{c}_{j}}\textrm{span}([\mathbf{H}^{m}_{j,j(1)}\mathbf{V}^{m}_{j(1)} \cdots \mathbf{H}^{m}_{j,j(M)}\mathbf{V}^{m}_{j(M)}])\right) \hspace{5mm} \\ \nonumber
&=& \hspace{-3mm} \textrm{dim} \Big([\mathbf{H}^{m}_{j,1(1)}\mathbf{V}^{m}_{1(1)} \cdots \mathbf{H}^{m}_{j,K(M)}\mathbf{V}^{m}_{K(M)}]\Big) \\ \nonumber
& &\hspace{-3mm}  - \textrm{dim} \Big([\mathbf{H}^{m}_{j,1(1)}\mathbf{V}^{m}_{1(1)} \cdots \mathbf{H}^{m}_{j,j-1(M)}\mathbf{V}^{m}_{j-1(M)} \\ \nonumber
& &\hspace{12mm}\mathbf{H}^{m}_{j,j+1(1)}\mathbf{V}^{m}_{j+1(1)} \cdots \mathbf{H}^{m}_{j,K(M)}\mathbf{V}^{m}_{K(M)}]\Big) \\ \nonumber
&\leq& \hspace{-3mm} m - \textrm{dim} (\mathcal{I}_{j}).
\end{eqnarray}
According to Lemma 3, we assume that all alignment sets consist of $n$ transmit symbols. These transmit symbols in the same alignment set are aligned at their common unintended receivers, while each of those symbols occupies an independent signal subspace at all of their corresponding receivers.
Since $n$ transmit symbols are aligned in the 1-dimensional signal subspace at their common unintended receivers, $(n-1)$ dimensions occupied by the interference signals are diminished.
If we denote the dimension of the aligned interfering signal subspaces occupied by a set of $n$ transmit 
symbols from transmit antennas $p_{1}(q_{1}),\ldots,p_{n}(q_{n})$ at their common unintended receivers by $d_{I}(p_{1}(q_{1}),\ldots ,p_{n}(q_{n}))$, the dimension of the signal subspace occupied by the interfering signals at $\mathcal{R}_{j}$ can be calculated by subtracting the dimension of the aligned interfering signal subspaces as
\begin{eqnarray}
\label{eq:I_j}
\textrm{dim} (\mathcal{I}_{j}) &=& \sum\limits_{k=1, k\neq j}^{K} d_{k}(m) - (n-1) \sum\limits_{\substack{p_{1}(q_{1}) \in \mathcal{B}, \\ p_{1} \neq j}} \\ \nonumber
& & \cdots 
\sum\limits_{\substack{p_{n}(q_{n}) \in \mathcal{B}, \\ p_{n} \neq j, \\ p_{n}(q_{n}) > p_{n-1}(q_{n-1})}}
d_{I}(p_{1}(q_{1}),\ldots ,p_{n}(q_{n})),
\end{eqnarray}
where $p_{n}(q_{n}) > p_{n-1}(q_{n-1})$ means that the
transmit antenna $p_{n}(q_{n})$ is the latter one than the transmit antenna $p_{n-1}(q_{n-1})$ in $\mathcal{B}$ when $\mathcal{B}$ has an order for elements from $1(1)$ to $K(M)$ as $\{1(1), \ldots, 1(M), 2(1), \ldots, K(M)\}$.
For $\mathcal{R}_{j}$, $d_{I}(p_{1}(q_{1}),\ldots ,p_{n}(q_{n}))$ is equal to $\textrm{dim}(\mathbf{H}^{m}_{j,p_{1}(q_{1})}\mathbf{V}^{m}_{p_{1}(q_{1})}\cap\cdots\cap \mathbf{H}^{m}_{j,p_{n}(q_{n})}\mathbf{V}^{m}_{p_{n}(q_{n})})$.
By rearranging (\ref{eq:lemma3}) and (\ref{eq:I_j}), the total dimension of signal subspaces over $m$ channel uses at $\mathcal{R}_{j}$ is
\begin{eqnarray}
\label{eq:dk} 
&& \hspace{-5mm} \sum\limits_{j=1}^{K}d_{j}(m) - (n-1) \sum\limits_{\substack{p_{1}(q_{1}) \in \mathcal{B}, \\ p_{1} \neq j}}\\ \nonumber
&& \hspace{5mm}  \cdots 
\sum\limits_{\substack{p_{n}(q_{n}) \in \mathcal{B}, \\ p_{n} \neq j, \\ p_{n}(q_{n}) > p_{n-1}(q_{n-1})}}
d_{I}(p_{1}(q_{1}),\ldots ,p_{n}(q_{n})) \leq m.
\end{eqnarray}
The best strategy to compose the alignment set of $n$ transmit symbols is to reduce the number of their corresponding receivers so as to be aligned in as many common unintended receivers as possible. 
For the $n$ transmit symbols from $n$ transmit antennas, the minimum number of their corresponding receivers is $\left\lceil\frac{n}{M}\right\rceil$ with $n$ transmit antennas from the $\left\lceil\frac{n}{M}\right\rceil$ transmitters, thereby a set of $n$ transmit symbols can be aligned in the interference signal subspace at $K-\left\lceil\frac{n}{M}\right\rceil$ unintended receivers.
By summing up (\ref{eq:dk}) over all the receivers,
\begin{eqnarray}
\label{eq:dall}
&& \hspace{-10mm} K \sum\limits_{j=1}^{K}d_{j}(m) - (n-1)(K-\left\lceil\frac{n}{M}\right\rceil) \sum\limits_{p_{1}(q_{1}) \in \mathcal{B}} \\ \nonumber
&&  \cdots
\sum\limits_{\substack{p_{n}(q_{n}) \in \mathcal{B}, \\ p_{n}(q_{n}) > p_{n-1}(q_{n-1})}}
d_{I}(p_{1}(q_{1}),\ldots ,p_{n}(q_{n})) \leq Km.
\end{eqnarray}
Meanwhile, according to Lemma 2, recall that if a transmit symbol is aligned with other $(n-1)$ symbols at their common unintended receivers, it cannot be aligned with the other set of $(n-1)$ symbols. This fact can be represented by the inequality as
\begin{eqnarray}
\label{eq:dlemma}
&&\hspace{-10mm}\sum\limits_{\substack{p_{1}(q_{1}) \in \mathcal{B}, \\ p_{1}(q_{1}) \neq i(a)}}\cdots
\sum\limits_{\substack{p_{n-1}(q_{n-1}) \in \mathcal{B}, \\ p_{n-1}(q_{n-1}) \neq i(a), \\ p_{n-1}(q_{n-1})> p_{n-2}(q_{n-2})}} \\ \nonumber
&& \hspace{5mm} d_{I}(i(a), p_{1}(q_{1}),\ldots ,p_{n-1}(q_{n-1})) \leq d_{i(a)}(m),
\end{eqnarray}
where $d_{i(a)}(m)$ denotes the dimension of the independent signal subspaces occupied by the transmit symbols from transmit antenna $i(a)$ at $\mathcal{R}_{i}$ over $m$ channel uses for $i(a) \in \mathcal{B}$, i.e., $\sum\nolimits_{a=1}^{M} d_{i(a)}(m)=d_{i}(m)$. From the fact that $KM\cdot\binom{KM-1}{n-1}$ is equal to $n\cdot\binom{KM}{n}$, we can derive the equality as
\begin{eqnarray}
\label{eq:dcom}
&& \hspace{-6mm} \sum\limits_{i(a) \in \mathcal{B}}\Big(\sum\limits_{\substack{p_{1}(q_{1}) \in \mathcal{B}, \\ p_{1}(q_{1}) \neq i(a)}} \cdots \\ \nonumber
&& \hspace{-5mm} \sum\limits_{\substack{p_{n-1}(q_{n-1}) \in \mathcal{B}, \\ p_{n-1}(q_{n-1}) \neq i(a), \\ p_{n-1}(q_{n-1}) > p_{n-2}(q_{n-2})}}
d_{I}(i(a), p_{1}(q_{1}),\ldots ,p_{n-1}(q_{n-1})) \Big) \hspace{8mm} \\ \nonumber 
&& \hspace{-5mm} = n \cdot \sum\limits_{p_{1}(q_{1}) \in \mathcal{B}} \cdots
\sum\limits_{\substack{p_{n}(q_{n}) \in \mathcal{B}, \\ p_{n}(q_{n}) > p_{n-1}(q_{n-1})}} 
d_{I}(p_{1}(q_{1}),\ldots ,p_{n}(q_{n})). 
\end{eqnarray}
The left-hand side represents the dimension of the interference signal subspace occupied by the alignment sets including the transmit symbol from transmit antenna $i(a)$, while the right-hand side represents the dimension of the interference signal subspace occupied by all the alignment sets, which does not separately consider the alignment sets including the transmit symbol from transmit antenna $i(a)$.
To take (\ref{eq:dlemma}) and (\ref{eq:dcom}) into account, (\ref{eq:dall}) changes to
\begin{eqnarray}
\label{eq:dsum}
K \sum\limits_{j=1}^{K}d_{j}(m) -
\frac{n-1}{n}\big(K-\left\lceil\frac{n}{M}\right\rceil \big)\sum\limits_{j=1}^{K}d_{j}(m) \leq Km.
\end{eqnarray}
Thus, we have
\begin{eqnarray}
\frac{1}{m} \sum\limits_{j=1}^{K}d_{j}(m) \leq 
\frac{nK}{K+\left\lceil\frac{n}{M}\right\rceil(n-1)}.
\label{eq:DoF}
\end{eqnarray}
This formula represents the linear sum DoF upper-bound when $n$ preset modes among $N$ are used.

Subsequently, the number of transmit symbols in an alignment set, $n$, should be determined to maximize 
the linear sum DoF upper-bound (\ref{eq:DoF}).
To be separated at their intended receivers, $n$ should not be greater than $N$ since receivers have at most $N$ 
independent channel states.
We define the LDoF function depending on $n$ as
\begin{eqnarray}
D(n)=\frac{nK}{K+\left\lceil\frac{n}{M}\right\rceil(n-1)}, \,\,\,\,\,\,\,\, n\leq N,
\end{eqnarray}
where $M$, $N$, and $K$ are given. Suppose that receivers have enough preset modes to align interfering signals, i.e., $N \geq MK$.
As it is observed in the LDoF function as
\begin{eqnarray}
\,\, D(M(\gamma -1)+\beta) < D(M\gamma), \,\,\,\,\, 0<\beta<M, \,\,\, \gamma , \beta \in \mathbb{N}
\label{eq:dbeta}
\end{eqnarray}
since the transmit symbols are aligned with a greater number of interference symbols at the same number of the unintended receivers when $n$ is a multiple of $M$ than the number of interference symbols when $n$ is not a multiple of $M$.
Thus, let $n$ be equal to $M\gamma$. Then, the LDoF function is
\begin{eqnarray}
D(M\gamma)=\frac{M\gamma K}{K+\gamma(M\gamma-1)}.
\end{eqnarray}
To maximize $D(M\gamma)$, we need to find $\gamma$ which minimizes $MK/D(M\gamma)=M\gamma-1+K/\gamma$.
By the inequality of arithmetic and geometric means,
\begin{eqnarray}
g(\gamma)=M\gamma+\frac{K}{\gamma},
\end{eqnarray}
is minimized when $\gamma=\sqrt\frac{K}{M}$. Since $\gamma$ is an integer and $g(\gamma)$ is a convex function, 
either $\left\lfloor\sqrt\frac{K}{M}\right\rfloor$ or $\left\lceil\sqrt\frac{K}{M}\right\rceil$ is a solution which 
minimizes $g(\gamma)$. We refer to it as $\Gamma_{opt}$. Therefore, the linear sum DoF upper-bound is maximized by $n^{*}$ as (\ref{eq:n}) 
when $N \geq M \cdot \left\lceil\sqrt{\frac{K}{M}}\right\rceil$.

Let us explore the case where $N=M\Gamma+\alpha$ is smaller than $M \cdot \left\lceil\sqrt{\frac{K}{M}}\right\rceil$.
Because $g(\gamma)$ is a convex function, 
\begin{eqnarray}
\,\,\,\,\,\,\,\, D(M(\gamma-1)) < D(M\gamma), \,\,\,\,\,\,\,\, \gamma < \left\lceil\sqrt{\frac{K}{M}}\right\rceil,
\label{eq:dgamma}
\end{eqnarray}
By (\ref{eq:dbeta}) and (\ref{eq:dgamma}),
\begin{eqnarray}
\,\,\,\,\,\,\,\,\,\,\,\,\,\,\,\, D(n) < D(M\Gamma), \,\,\,\,\,\,\,\, n < M\Gamma, \Gamma < \left\lceil\sqrt{\frac{K}{M}}\right\rceil.
\label{eq:dGamma}
\end{eqnarray}
Meanwhile, it is easily shown that
\begin{eqnarray}
\,\,\,\,\,\,\,\, D(M\gamma+\beta_{1}) < D(M\gamma+\beta_{2}), \,\,\,\, 0 < \beta_{1} < \beta_{2}<M,
\label{eq:dBeta}
\end{eqnarray}
since $D(M\gamma+\beta_{2})$ shows the upper-bound of the case where the alignment sets include a greater number of transmit symbols than $D(M\gamma+\beta_{1})$, while the alignment sets are aligned at the same number of unintended receivers.
Consequently, $n^{*}$ should be $M\Gamma$ or $M\Gamma+\alpha$ by considering the inequalities (\ref{eq:dGamma}) and (\ref{eq:dBeta}).
$D(M\Gamma+\alpha)$ shows the upper-bound when all the preset modes are used to align interference signals, while $D(M\Gamma)$ shows the upper-bound when only $M\Gamma$ preset modes are used to align the alignment set at one more unintended receivers than the previous case.
Thus, we conclude that $n^{*}$ is determined as (\ref{eq:n}) by comparing $D(M\Gamma)$ and $D(M\Gamma+\alpha)$,
i.e., $D(M\Gamma+\alpha)$ is greater than $D(M\Gamma)$ if $\alpha > \frac{N(M\Gamma-1)}{K-\Gamma-1}$.

\begin{figure}[t]
    \centerline{\includegraphics[width=9.0cm]{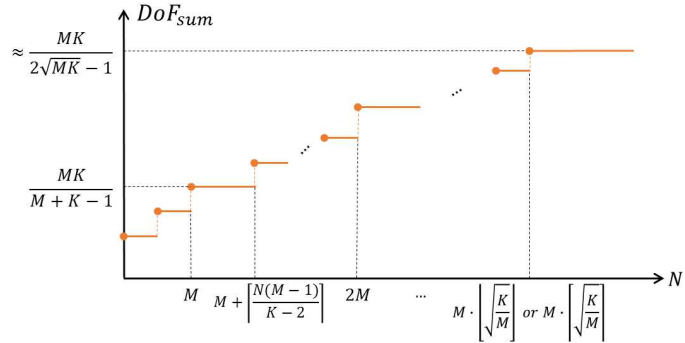}}
    \caption{Linear sum DoF upper-bound characterization for $K$-user $M \times 1$ interference channels with blind interference alignment through reconfigurable antenna switching.}
    \label{DoF}
    \vspace{-3mm}
\end{figure}

\subsection{Discussion}
\label{sec_dis}
In this section, we discuss the linear sum DoF upper-bound tendency as $N$ increases. We also show the linear sum DoF upper-bound for the $K$-user SISO IC, and compare it to the previous DoF characterization results in \cite{IC3}-\cite{IC4} and finite state compound wireless network scenarios \cite{Compound}. 

Fig. \ref{DoF} shows the linear sum DoF upper-bound characterization for the $K$-user $M \times 1$ IC with reconfigurable antenna switching as $N=M\Gamma+\alpha$ increases according to the result of Theorem 1.
When $N<M$ $(\Gamma=0)$, $n^{*}$, which maximizes the upper-bound is $N$ since $\alpha$ is always greater than $\frac{N(M\cdot 0-1)}{K-0-1}=\frac{-N}{K-1}$. 
If $N=M$, the upper-bound is $\frac{MK}{M+K-1}$ that is the same result as \cite{IC2}.
For the $M\gamma < N < M(\gamma+1)$ region, there is a distinct separation of linear sum DoF upper-bound increase tendency
by the criteria, $\alpha = \left\lceil\frac{N(M\gamma-1)}{K-\gamma-1}\right\rceil$.
When $\alpha < \left\lceil\frac{N(M\gamma-1)}{K-\gamma-1}\right\rceil$, the upper-bound 
does not increase compared to the  
$N=M\gamma$ case since $n^{*}$ is still $M\gamma$ as in (\ref{eq:n}).
On the other hand, when $\alpha \geq \left\lceil\frac{N(M\gamma-1)}{K-\gamma-1}\right\rceil$, the upper-bound 
can be enhanced by utilizing all the preset modes since $n^{*}=M\gamma+\alpha$.
This tendency continues until
\begin{eqnarray}
\gamma = \Gamma_{opt} =\underset{\gamma = \left\lfloor\sqrt{\frac{K}{M}}\right\rfloor, \left\lceil\sqrt{\frac{K}{M}}\right\rceil}
{\mathrm{argmin}} M\gamma+\frac{K}{\gamma},
\end{eqnarray}
according to (\ref{eq:n}).
Even when $N$ increases beyond $M\cdot\Gamma_{opt}$, the upper-bound stays the same.
The maximum sum DoF upper-bound with sufficiently large $N$ is close to $\frac{MK}{2\sqrt{MK}-1}$ by assuming that $\Gamma_{opt}$ is close to $\sqrt{\frac{K}{M}}$.

\textbf{Corollary 2} \cite{SISO}:
For the $K$-user SISO IC with $N$ preset modes at the receivers, i.e., $(1,N,K)$-IC, 
the linear sum DoF is upper-bounded by
\begin{eqnarray}
\textrm{LDoF}_{\textrm{sum}} \leq \frac{n^{*}K}{K+n^{*}(n^{*}-1)},
\end{eqnarray}
where
\begin{eqnarray}
n^{*}=\left\{\begin{array}{cc} 
N & N < \left\lceil\sqrt{K}\right\rceil,\\
\underset{n = \left\lfloor\sqrt{K}\right\rfloor, \left\lceil\sqrt{K}\right\rceil}
{\mathrm{argmin}} n+\frac{K}{n} & N \geq \left\lceil\sqrt{K}\right\rceil. \end{array} \right.
\end{eqnarray}

We simplify Theorem 1 for the SISO IC case by setting $M=1$. It is observed that our result includes
previous work \cite{IC3}-\cite{IC4}. Moreover, we prove that preset modes more than 2 are also effective for SISO IC to 
increase the linear sum DoF. According to the result, 3 preset modes increase the linear sum DoF when $K \geq 7$
and 4 preset modes are effective when $K \geq 13$.

\textbf{Remark 1} [Comparison with finite state compound wireless networks]:
The considered system model in this paper is similar to finite state compound wireless networks in \cite{Compound} since 
the channel state should be determined among elements of a finite set for both of the two system models.
According to Corollary 2, the linear sum DoF upper-bound for $K$-user SISO IC is close to $\frac{K}{2\sqrt{K}-1}$ when $\left\lfloor\sqrt{K}\right\rfloor$ and $\left\lceil\sqrt{K}\right\rceil$ are close to $\sqrt{K}$. 
However, in \cite{Compound}, the $K$-user finite state compound IC achieves a total of $K/2$ DoF even if CSIT is not available. 
The BIA scheme introduced in \cite{blind}-\cite{blind2} also achieves the full $K/2$ DoF
for the $K$-user SISO IC by exploiting specific channel correlations. In this scheme, the channel state of the interfering link is assumed 
to remain constant, while the channel state of the direct links varies.
\footnote{This assumption could be feasible, although it is an unusual scenario for 
practical communication systems,
when all the direct links are time-selective and all the interfering links are frequency-selective, or vice versa.}
This result is completely different from the result of the BC scenario
that a total of $\frac{MK}{M+K-1}$ DoF is achievable for the $K$-user $M \times 1$ MISO BC scheme with reconfigurable antennas 
at the receivers, which coincides with the achievable sum DoF of finite state compound BC.  
It is interesting to note that the upper-bound for IC with a reconfigurable antenna cannot meet the sum DoF for finite 
state compound IC, which is because the channel states' changing pattern towards the same receiver are always equivalent.
When the direct link and the interfering link towards the same receiver have the same changing pattern, 
interfering signals can be aligned at a limited number of receivers as shown in Lemma 2.
On the other hand, it causes no influence on the BC scenario since the desired symbol and the interfering symbol are transmitted via the same link,
thus the direct link and the interfering link (the same as the direct link) always have the same changing pattern in every system model as well as the reconfigurable antenna switching scheme.
Consequently, it can be said that the linear sum DoF upper-bound for the $K$-user SISO IC increases
from 1 \cite{Huang}-\cite{Vaze} to $\frac{K}{2\sqrt{K}-1}$ in the absence of CSIT by reconfigurable antenna switching,
however, the full $K/2$ DoF is not achievable in this scheme due to the inherent feature that the channel states of the links towards the same receiver inevitably have the identical changing pattern.

\section{DoF Achievability}
\label{sec_ach}
We dedicate this section to show how the linear sum DoF upper-bound explained in Section \ref{sec_DoF} 
can be achieved by each user's preset mode pattern and beamforming vectors.
We notate $\frac{n^{*}}{M}$ and $\frac{MK}{n^{*}}$ as  $R_{1}$ and $R_{2}$, respectively, and classify three scenarios whether
$R_{1}$ and $R_{2}$ are integer or not: 1) $R_{1}$ and $R_{2}$ are integers
2) $R_{1}$ is an integer, but $R_{2}$ is not an integer 3) $R_{1}$ is not an integer.
We show that the linear sum DoF upper-bound is achievable for the first case by proving Theorem 2. For the other cases, we give some guidelines for an achievable scheme how to make an alignment set with the transmit symbols.

\subsection{$R_{1}$ and $R_{2}$ are integers. ($n^{*}$ is a multiple of $M$ and $MK$ is divisible by $n^{*}$.)}
\label{sec_ach_1}
In this case, we modify the existing achievable scheme for the $K$-user MISO IC \cite{IC2} to our scenario.
In Theorem 2, we show that if $R_{1}$ and $R_{2}$ are integers where $R_{1}=\frac{n^{*}}{M}$, $R_{2}=\frac{MK}{n^{*}}$, and
$n^{*}$ is determined by (\ref{eq:n}), the linear sum DoF upper-bound (\ref{eq:DoF_n}) is achievable for the 
$K$-user $M \times 1$ MISO IC with $N$ preset modes at each receiver when beamforming vectors and 
preset mode patterns are constructed as $R_{2}$-user $n^{*} \times 1$ MISO IC according to \cite{IC2}.

\emph{Proof of Theorem 2}:
According to Lemma 2, the transmit symbols in an alignment set cannot be aligned at 
their corresponding receivers which desire one of the transmit symbols.
When we group $n^{*}$ transmit symbols to be aligned at their common unintended receivers, they have  independent signal subspaces at their corresponding receivers. It means that it is sufficient for the corresponding receivers to have the same preset mode pattern to decode those transmit symbols in an alignment set.
First, we partition the K transmitter-receiver pairs into $R_2$ groups of size $R_1$.
Note that the total number of antennas for each group (consisting of $R_1$ users) is 
$n^{*} = M R_{1}$.
We then design the beamforming vectors and the preset mode patterns for an equivalent
$R_{2}$-user $n^{*} \times 1$ MISO IC by the method of \cite{IC2}, where 
each one of $R_{2}$ transmitter-receiver pairs corresponds to a group of
$R_1$ transmitter-receiver pairs in our system model.
Note also that the preset mode pattern for each receiver in the equivalent system
is shared by the receivers of its corresponding group in the original system.
Specifically, based on the achievable scheme in \cite{IC2}, each transmit antenna sends $(n^{*}-1)^{R_{2}-1}$ transmit symbols to achieve the linear sum DoF upper-bound, and the receivers can decode the transmit symbols in the alignment sets, which include their desired symbols.
The beamforming vectors for each transmitter coincide with those of the IC scheme in \cite{IC2} as
\begin{eqnarray}
\mathbf{V}^{m}_{i(a)} = \widehat{\mathbf{V}}^{m}_{i^{'} ((i-1)M+a-(i^{'}-1)n^{*})},
\end{eqnarray}
where $\widehat{\mathbf{V}}^{m}_{i^{'}(b)}$ denotes the beamforming vector of the $b^{th}$ transmit antenna of $\mathcal{T}_{i^{'}}$ for $R_{2}$-user $n^{*} \times 1$ MISO IC and $i^{'}=\left\lceil \frac{(i-1)M+a}{n^{*}} \right\rceil$.
Likewise, the preset mode patterns also coincide with those of the IC scheme in \cite{IC2} as
\begin{eqnarray}
\mathbf{L}^{m}_{j}=\widehat{\mathbf{L}}^{m}_{\left\lceil j/R_{1} \right\rceil}.
\end{eqnarray}
where $\widehat{\mathbf{L}}^{m}_{j^{'}}$ denotes the preset mode pattern of $\mathcal{R}_{j^{'}}$ for the $R_{2}$-user $n^{*} \times 1$ MISO IC.
By this achievable scheme, $R_{1}$ users achieve a total of $\frac{n^{*}}{n^{*}+R_{2}-1}$ DoF,
so that the achievable linear sum DoF is
\begin{eqnarray}
\textrm{LDoF}_{\textrm{sum}}&=&\frac{n^{*}\cdot R_{2}}{n^{*}+R_{2}-1} \\ \nonumber
&=&\frac{n^{*}\cdot \frac{MK}{n^{*}}}{n^{*}+\frac{MK}{n^{*}}-1} \\ \nonumber
&=&\frac{n^{*}\cdot MK}{MK+n^{*}(n^{*}-1)} \\ \nonumber
&=&\frac{n^{*}K}{K+\frac{n^{*}}{M}(n^{*}-1)}.
\end{eqnarray}
This result is equal to the linear sum DoF upper-bound in (\ref{eq:DoF_n}). 

\emph{Example 3}:
Consider the case of $(1,2,4)$-IC.
In this case, $n^{*}$ is equal to 2, thereby $n^{*}$ is a multiple of $M=2$ and $MK=4$ is divisible by $n^{*}$,
i.e., $R_{1}$ and $R_{2}$ are integers.
When two users are grouped to act as a single user, they can be seen as the 2-user $2 \times 1$ MISO IC.
The transmit beamforming vectors for each transmitter during 3-symbol extension are
\begin{eqnarray}
\mathbf{v}^{3}_{1,1} = \mathbf{v}^{3}_{2,1} = [1 \hspace{1mm} 1 \hspace{1mm} 0]^{T}, \hspace{5mm}
\mathbf{v}^{3}_{3,1} = \mathbf{v}^{3}_{4,1} = [1 \hspace{1mm} 0 \hspace{1mm} 1]^{T},
\end{eqnarray}
when each user transmits one data symbol.
The preset mode pattern for each receiver during 3-symbol extension is
\begin{eqnarray}
\mathbf{L}^{3}_{1} = \mathbf{L}^{3}_{2} = [1 \hspace{1mm} 2 \hspace{1mm} 1], \hspace{5mm}
\mathbf{L}^{3}_{3} = \mathbf{L}^{3}_{4} = [1 \hspace{1mm} 1 \hspace{1mm} 2],
\end{eqnarray}
In this case, the received signal during 3-symbol extension at $\mathcal{R}_{1}$ is
\begin{eqnarray}
\mathbf{y}^{3}_{1} \hspace{-1mm} = \hspace{-1mm} \left[ \hspace{-1mm} \begin{array}{cccc}
h_{1,1}(1) & h_{1,2}(1) & h_{1,3}(1) & h_{1,4}(1) \\
h_{1,1}(2) & h_{1,2}(2) & 0 & 0 \\
0 & 0 & h_{1,3}(1) & h_{1,4}(1) \end{array} \hspace{-1mm} \right] \hspace{-2mm}
\left[\hspace{-1mm} \begin{array}{c}
s_{1,1} \\ s_{2,1} \\ s_{3,1} \\ s_{4,1} \end{array} \hspace{-1mm} \right]
\hspace{-1mm} + \mathbf{z}^{3}_{1}.
\end{eqnarray}
$\mathcal{R}_{1}$ can decode $s_{1,1}$ and $s_{2,1}$ by subtracting the received signal at time 3 from the signal at time 1.
Since $s_{1,1}$ is a desired symbol at $\mathcal{R}_{1}$, user 1 achieves 1 DoF during 3-symbol extension.
Because $\mathcal{R}_{2}$ has the same preset mode pattern as $\mathcal{R}_{1}$, it also achieves $s_{1,1}$ and $s_{2,1}$ 
in the same way. On the other hand, $\mathcal{R}_{[3:4]}$ achieve $s_{3,1}$ and $s_{4,1}$ 
by subtracting the received signal at time 2 from the signal at time 1.
Consequently, a total of $4/3$ DoF is achievable by this scheme. It coincides with the result of Theorem 1.

\subsection{$R_{1}$ is an integer, but $R_{2}$ is not an integer. ($n^{*}$ is a multiple of $M$ and $MK$ is not divisible by $n^{*}$.)}
\label{sec_ach_2}
In this case, a new achievable scheme is required since the existing achievable scheme cannot be directly applied.
According to the proof of the LDoF converse, 
all the transmit symbols need to be included in the alignment sets in order that an alignment set has $n^{*}$ transmit symbols to achieve the linear sum DoF upper-bound.
If not, there is DoF loss compared to the upper-bound since some transmit symbols are not optimally aligned
at their unintended receivers.
Therefore, users need to increase the number of their transmit symbols $\eta$ times in order that the number of transmit symbols 
from all transmitters is divisible by the cardinality of the alignment sets, i.e., $\eta MK$ should be divisible by $n^{*}$.

\begin{figure}[t]
    \centerline{\includegraphics[width=8.5cm]{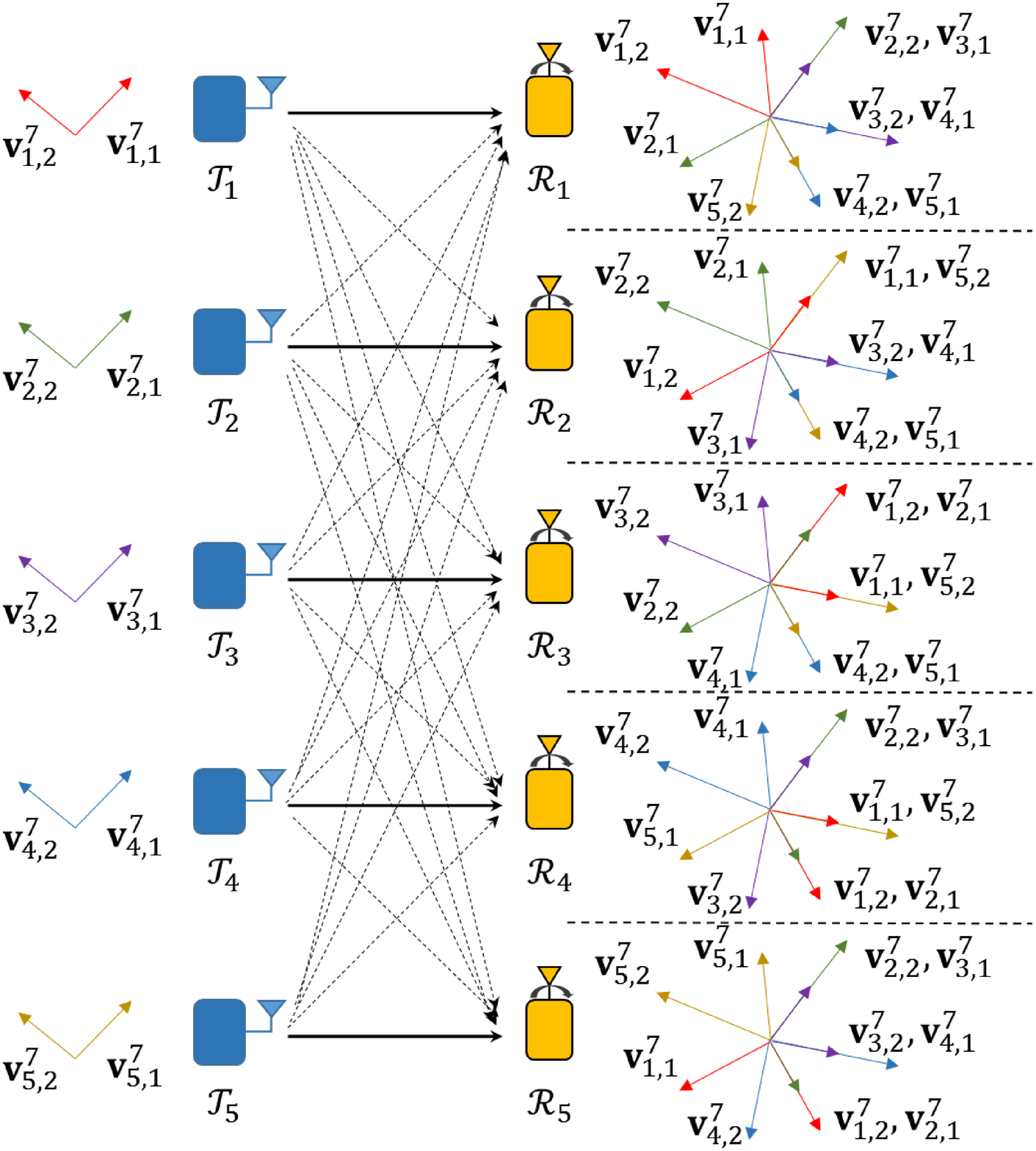}}
    \caption{Description for the 5-user SISO IC with 2 preset modes, i.e., $(1,2,5)$-IC.}
    \label{MNK125}
    \vspace{-3mm}
\end{figure}

\emph{Example 4}:
Let us introduce the case of $(1,2,5)$-IC, where $n^{*}=2$.
Each user needs to send 2 data symbols to achieve the linear sum DoF upper-bound, so that 10 symbols are grouped to 5 alignment sets.
The alignment sets are constructed as
\begin{eqnarray}
\begin{array}{cc}
\mathcal{A}_{\{1,2\}}=\{\mathbf{v}^{7}_{1,2}, \mathbf{v}^{7}_{2,1}\}, &
\mathcal{A}_{\{2,3\}}=\{\mathbf{v}^{7}_{2,2}, \mathbf{v}^{7}_{3,1}\}, \\
\mathcal{A}_{\{3,4\}}=\{\mathbf{v}^{7}_{3,2}, \mathbf{v}^{7}_{4,1}\}, &
\mathcal{A}_{\{4,5\}}=\{\mathbf{v}^{7}_{4,2}, \mathbf{v}^{7}_{5,1}\}, \\ 
\mathcal{A}_{\{1,5\}}=\{\mathbf{v}^{7}_{5,2}, \mathbf{v}^{7}_{1,1}\}, & 
\end{array}
\end{eqnarray}
which is described in Fig. \ref{MNK125}.
The beamforming vectors for each data symbols during 7-symbol extension can be determined as 
\begin{eqnarray}
&&\mathbf{v}^{7}_{1,2} = \mathbf{v}^{7}_{2,1} = [1 \hspace{1mm} 1 \hspace{1mm} 1 \hspace{1mm} 0 \hspace{1mm} 0 \hspace{1mm} 0 \hspace{1mm} 0]^{T}, \\ \nonumber
&&\mathbf{v}^{7}_{2,2} = \mathbf{v}^{7}_{3,1} = [0 \hspace{1mm} 1 \hspace{1mm} 0 \hspace{1mm} 0 \hspace{1mm} 1 \hspace{1mm} 0 \hspace{1mm} 0]^{T}, \\ \nonumber
&&\mathbf{v}^{7}_{3,2} = \mathbf{v}^{7}_{4,1} = [0 \hspace{1mm} 0 \hspace{1mm} 1 \hspace{1mm} 0 \hspace{1mm} 0 \hspace{1mm} 1 \hspace{1mm} 0]^{T}, \\ \nonumber
&&\mathbf{v}^{7}_{4,2} = \mathbf{v}^{7}_{5,1} = [0 \hspace{1mm} 0 \hspace{1mm} 0 \hspace{1mm} 0 \hspace{1mm} 1 \hspace{1mm} 0 \hspace{1mm} 1]^{T}, \\ \nonumber
&&\mathbf{v}^{7}_{5,2} = \mathbf{v}^{7}_{1,1} = [1 \hspace{1mm} 0 \hspace{1mm} 0 \hspace{1mm} 1 \hspace{1mm} 0 \hspace{1mm} 0 \hspace{1mm} 0]^{T}.
\end{eqnarray}
The preset mode pattern corresponding to the beamforming vectors is
\begin{eqnarray}
\begin{array}{cc}
\mathbf{L}^{7}_{1}=[1 \hspace{1mm} 1 \hspace{1mm} 2 \hspace{1mm} 2 \hspace{1mm} 1 \hspace{1mm} 2 \hspace{1mm} 1], &
\mathbf{L}^{7}_{2}=[1 \hspace{1mm} 2 \hspace{1mm} 2 \hspace{1mm} 1 \hspace{1mm} 1 \hspace{1mm} 2 \hspace{1mm} 1], \\
\mathbf{L}^{7}_{3}=[1 \hspace{1mm} 1 \hspace{1mm} 1 \hspace{1mm} 1 \hspace{1mm} 2 \hspace{1mm} 2 \hspace{1mm} 2], &
\mathbf{L}^{7}_{4}=[1 \hspace{1mm} 1 \hspace{1mm} 1 \hspace{1mm} 1 \hspace{1mm} 1 \hspace{1mm} 2 \hspace{1mm} 2], \\
\mathbf{L}^{7}_{5}=[1 \hspace{1mm} 1 \hspace{1mm} 1 \hspace{1mm} 2 \hspace{1mm} 1 \hspace{1mm} 1 \hspace{1mm} 2]. &
\end{array}
\end{eqnarray}
The received signal at $\mathcal{R}_{1}$ is (\ref{eq:512}).
We can see that the alignment sets which have no desired signal for $\mathcal{R}_{1}$, i.e., $\mathcal{A}_{\{2,3\}}$, 
$\mathcal{A}_{\{3,4\}}$, and $\mathcal{A}_{\{4,5\}}$ are aligned in an 1-dimensional signal subspace, while 
the desired signals for $\mathcal{R}_{1}$ and the grouped signal with them have independent signal subspaces.
It can be verified by showing that $\mathbf{R}_{1}$ in (\ref{eq:R1}),
whose columns represent the vectors carrying all the desired symbols and one symbol out of the interfering symbols of each alignment set, has full rank.
We also verify that other users have the same DoF as user 1 by confirming $\mathbf{R}_{2}, \ldots, \mathbf{R}_{5}$ have 
full rank.
By this scheme, each user achieves 2 DoF over 7-symbol extension, so that the linear sum DoF is $10/7$, 
which is the same as the upper-bound.
\begin{figure*}[ht]
{\small\begin{eqnarray} \label{eq:512}
\mathbf{y}^{7}_{1}= \left[ \hspace{-1mm} \begin{array}{cc}
h_{1,1}(1) & h_{1,1}(1) \\ 0 & h_{1,1}(1) \\ 0 & h_{1,1}(2) \\ h_{1,1}(2) & 0 \\ 
0 & 0 \\ 0 & 0 \\ 0 & 0 \end{array} \hspace{-1mm} \right] \hspace{-1mm}
\left[ \hspace{-1mm} \begin{array}{c} s_{1,1} \\ s_{1,2} \end{array} \hspace{-1mm} \right] +
\left[\hspace{-1mm} \begin{array}{cccccccc}
h_{1,2}(1) & 0 & 0 & 0 & 0 & 0 & 0 & h_{1,5}(1) \\
h_{1,2}(1) & h_{1,2}(1) & h_{1,3}(1) & 0 & 0 & 0 & 0 & 0 \\
h_{1,2}(2) & 0 & 0 & h_{1,3}(2) & h_{1,4}(2) & 0 & 0 & 0 \\
0 & 0 & 0 & 0 & 0 & 0 & 0 & h_{1,5}(2) \\
0 & h_{1,2}(1) & h_{1,3}(1) & 0 & 0 & h_{1,4}(1) & h_{1,5}(1) & 0 \\
0 & 0 & 0 & h_{1,3}(2) & h_{1,4}(2) & 0 & 0 & 0 \\
0 & 0 & 0 & 0 & 0 & h_{1,4}(1) & h_{1,5}(1) & 0 \end{array} \hspace{-1mm} \right] \hspace{-1mm}
\left[ \hspace{-1mm} \begin{array}{c} s_{2,1} \\ s_{2,2} \\ s_{3,1} \\ s_{3,2} \\
s_{4,1} \\ s_{4,2} \\ s_{5,1} \\ s_{5,2} \end{array} \hspace{-1mm} \right] + \mathbf{z}^{7}_{1}
\end{eqnarray}}
{\small\begin{eqnarray} \label{eq:R1}
\mathbf{R}_{1}\triangleq \left[ \begin{array}{ccccccc}
h_{1,1}(1) & h_{1,1}(1) & h_{1,2}(1) & 0 & 0 & 0 & h_{1,5}(1) \\
0 & h_{1,1}(1) & h_{1,2}(1) & h_{1,2}(1) & 0 & 0 & 0 \\
0 & h_{1,1}(2) & h_{1,2}(2) & 0 & h_{1,3}(2) & 0 & 0 \\
h_{1,1}(2) & 0 & 0 & 0 & 0 & 0 & h_{1,5}(2) \\
0 & 0 & 0 & h_{1,2}(1) & 0 & h_{1,4}(1) & 0 \\
0 & 0 & 0 & 0 & h_{1,3}(2) & 0 & 0 \\
0 & 0 & 0 & 0 & 0 & h_{1,4}(1) & 0 \end{array} \right],
\end{eqnarray}}
\hrulefill \vspace{-3mm}
\end{figure*}

\subsection{$R_{1}$ is not an integer. ($n^{*}$ is not a multiple of $M$.)}
\label{sec_ach_3}
In this case, the achievable scheme should be modified to equalize the number of symbol extensions between users.
If $n^{*}$ is not a multiple of $M$, the total signal dimensions at each receiver could be different. It causes different number of symbol extensions between users, which makes it difficult to design a proper achievable scheme. To make an appropriate beamforming vectors and preset mode patterns, users should increase the number of transmit symbols in order to equalize the number of symbol extensions between users.

\emph{Example 5}:
Consider a $(2,3,6)$-IC scenario where $n^{*}=3$. If each transmit antenna sends a single transmit symbol, transmit symbols could be included in the alignment set to be aligned at the unintended receivers as follows
\begin{eqnarray}
\begin{array}{c}
\mathcal{A}_{\{1,2\}}=\{\mathbf{v}^{m}_{1(1),1}, \mathbf{v}^{m}_{1(2),1}, \mathbf{v}^{m}_{2(1),1}\}, \\
\mathcal{A}_{\{2,3\}}=\{\mathbf{v}^{m}_{2(2),1}, \mathbf{v}^{m}_{3(1),1}, \mathbf{v}^{m}_{3(2),1}\}, \\ 
\mathcal{A}_{\{4,5\}}=\{\mathbf{v}^{m}_{4(1),1}, \mathbf{v}^{m}_{4(2),1}, \mathbf{v}^{m}_{5(1),1}\}, \\
\mathcal{A}_{\{5,6\}}=\{\mathbf{v}^{m}_{5(2),1}, \mathbf{v}^{m}_{6(1),1}, \mathbf{v}^{m}_{6(2),1}\}.
\end{array}
\end{eqnarray}
Each alignment set occupies an 1-dimensional signal subspace at the unintended receivers, while it occupies a 3-dimensional signal subspace at the corresponding receivers for each symbol to have an independent subspace.
In this situation, $\mathcal{R}_{1}$ needs 6-symbol extension since the alignment sets $\mathcal{A}_{\{2,3\}}$, $\mathcal{A}_{\{4,5\}}$, and $\mathcal{A}_{\{5,6\}}$ occupy an 1-dimensional subspace and $\mathcal{A}_{\{1,2\}}$ occupies a 3-dimensional subspace. On the other hand, $\mathcal{R}_{2}$ needs 8-symbol extension because $\mathcal{A}_{\{1,2\}}$ and $\mathcal{A}_{\{2,3\}}$ occupy a 3-dimensional subspace each. This is because $\mathcal{R}_{1}$ has the desired symbols (i.e. $\mathbf{v}^{m}_{1(1),1}, \mathbf{v}^{m}_{1(2),1}$) only in $\mathcal{A}_{\{1,2\}}$, while $\mathcal{R}_{2}$ has the desired symbols (i.e. $\mathbf{v}^{m}_{2(1),1}, \mathbf{v}^{m}_{2(2),1}$) in $\mathcal{A}_{\{1,2\}}$ and $\mathcal{A}_{\{2,3\}}$. To make all the receivers have the same number of symbol extensions, we suppose that each transmit antenna sends three transmit symbols. The transmit symbols could be included in the alignment set as 
\begin{eqnarray}
\begin{array}{c}
\mathcal{A}^{1}_{\{1,2\}}=\{\mathbf{v}^{20}_{(1)1,1}, \mathbf{v}^{20}_{(1)2,1}, \mathbf{v}^{20}_{(2)1,1}\}, \\
\mathcal{A}^{1}_{\{2,3\}}=\{\mathbf{v}^{20}_{(2)2,1}, \mathbf{v}^{20}_{(3)1,1}, \mathbf{v}^{20}_{(3)2,1}\}, \\
\mathcal{A}^{1}_{\{4,5\}}=\{\mathbf{v}^{20}_{(4)1,1}, \mathbf{v}^{20}_{(4)2,1}, \mathbf{v}^{20}_{(5)1,1}\}, \\
\mathcal{A}^{1}_{\{5,6\}}=\{\mathbf{v}^{20}_{(5)2,1}, \mathbf{v}^{20}_{(6)1,1}, \mathbf{v}^{20}_{(6)2,1}\}, \\ 
\mathcal{A}^{2}_{\{1,2\}}=\{\mathbf{v}^{20}_{(1)2,2}, \mathbf{v}^{20}_{(2)1,2}, \mathbf{v}^{20}_{(2)2,2}\}, \\
\mathcal{A}^{1}_{\{3,4\}}=\{\mathbf{v}^{20}_{(3)1,2}, \mathbf{v}^{20}_{(3)2,2}, \mathbf{v}^{20}_{(4)1,2}\}, \\ 
\mathcal{A}^{2}_{\{4,5\}}=\{\mathbf{v}^{20}_{(4)2,2}, \mathbf{v}^{20}_{(5)1,2}, \mathbf{v}^{20}_{(5)2,2}\}, \\
\mathcal{A}^{1}_{\{1,6\}}=\{\mathbf{v}^{20}_{(6)1,2}, \mathbf{v}^{20}_{(6)2,2}, \mathbf{v}^{20}_{(1)1,2}\}, \\
\mathcal{A}^{2}_{\{2,3\}}=\{\mathbf{v}^{20}_{(2)1,3}, \mathbf{v}^{20}_{(2)2,3}, \mathbf{v}^{20}_{(3)1,3}\},\\
\mathcal{A}^{2}_{\{3,4\}}=\{\mathbf{v}^{20}_{(3)2,3}, \mathbf{v}^{20}_{(4)1,3}, \mathbf{v}^{20}_{(4)2,3}\}, \\
\mathcal{A}^{2}_{\{5,6\}}=\{\mathbf{v}^{20}_{(5)1,3}, \mathbf{v}^{20}_{(5)2,3}, \mathbf{v}^{20}_{(6)1,3}\}, \\
\mathcal{A}^{2}_{\{1,6\}}=\{\mathbf{v}^{20}_{(6)2,3}, \mathbf{v}^{20}_{(1)1,3}, \mathbf{v}^{20}_{(1)2,3}\}.
\end{array}
\end{eqnarray}
In that case, all users have the desired signals in 4 alignment sets. Thus, they need 20-symbol extensions because 4 alignment sets occupy a 3-dimensional subspace each, while other 8 alignment sets occupy an 1-dimensional subspace. Therefore, each user achieves 6 DoF over 20-symbol extensions, thus the linear sum DoF is 9/5, which meets the upper-bound for the $(2,3,6)$-IC.
In this way of constructing alignment sets, a proper achievable scheme can be designed, so that users have the extension of the same number of symbols.

\textbf{Remark 2}:
If the considered ($M,N,K$)-IC scenario does not meet the two conditions, 1) $n^{*}$ is a multiple of $M$ and 2) $MK$ is divisible by $n^{*}$, the proposed achievable scheme in Theorem 2 cannot be applied.  
Moreover, to align interfering signals at the receiver in order to meet the linear sum DoF upper-bound, more transmit symbols need to be sent than introduced in Section \ref{sec_ach_2} and \ref{sec_ach_3} if $n^{*} > 2$.
As an example, for the $K$-user $M \times 1$ MISO IC scenario with $M$ preset modes \cite{IC2}, each transmitter sends $(M-1)^{K-1}$ transmit symbols to achieve a total of $\frac{MK}{M+K-1}$ DoF, which is the linear sum DoF upper-bound. We describe an alignment strategy where the number of transmit symbols for each transmit antenna in Section IV-B
and IV-C is sufficient to achieve the upper-bound.
Thus, we just suggest a guideline in Section \ref{sec_ach_2} and \ref{sec_ach_3} how the transmit symbols from each transmit antenna are aligned at the unintended receivers to meet the linear sum DoF upper-bound by introducing the concept of an alignment set.
It could be interesting future work to propose the general framework for ($M,N,K$)-IC
that can be applied even if the mentioned two conditions are not satisfied.


\section{Extension to cellular networks}
\label{sec_cell}

The linear sum upper-bound can be simply extended to fully connected interfering broadcast channel, i.e., downlink cellular networks,
with a similar approach inspired by Lemma 2.

\begin{figure}[t]
    \centerline{\includegraphics[width=8.0cm]{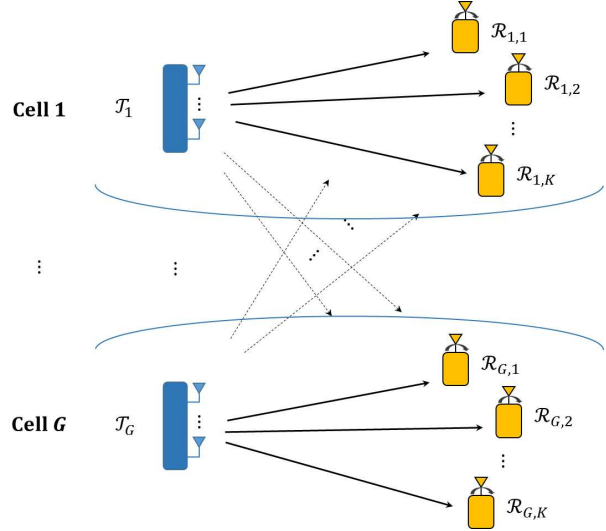}}
    \caption{$G$-cell $M\times1$ MISO fully connected interfering broadcast channels, where $K$ receivers exist in each cell.}
    \label{IBC}
    \vspace{-3mm}
\end{figure}

\textbf{Corollary 3}:
For $G$-cell $M\times1$ MISO fully connected interfering broadcast channels which are illustrated in Fig. \ref{IBC}, where $K$ receivers exist in each cell equipped with a single reconfigurable antenna with $N$ preset modes, the linear sum DoF is upper-bounded by
\begin{eqnarray}
\label{eq:DoF_IBC}
\textrm{LDoF}_{\textrm{sum}} \leq \frac{n^{*}KG}{KG+\left\lceil\frac{n^{*}}{M}\right\rceil(n^{*}-1)},
\end{eqnarray}
where $\overline{N}=\textrm{min}\{N, MG\}$ is expressed as $M\Gamma+\alpha$ ($0 \leq \alpha < M$),
and $n^{*}$ represents the optimal number of preset modes as
\begin{eqnarray} \label{eq:n_IBC}
n^{*}=\left\{\begin{array}{cc}
M\Gamma, & \overline{N} < M \cdot \left\lceil\sqrt{\frac{KG}{M}}\right\rceil, \alpha \leq \frac{\overline{N}(M\Gamma-1)}{KG-\Gamma-1} \\
\overline{N}, & \overline{N} < M \cdot \left\lceil\sqrt{\frac{KG}{M}}\right\rceil, \alpha > \frac{\overline{N}(M\Gamma-1)}{KG-\Gamma-1} \\
M \cdot \Gamma_{opt},  & \overline{N} \geq M \cdot \left\lceil\sqrt{\frac{KG}{M}}\right\rceil \end{array} \right.
\end{eqnarray}
where $\Gamma_{opt}=\underset{\gamma = \left\lfloor\sqrt{\frac{KG}{M}}\right\rfloor, \left\lceil\sqrt{\frac{KG}{M}}\right\rceil}
{\textrm{argmin}} M\gamma+\frac{KG}{\gamma}$.
\begin{proof}
It can be proved by a similar way as in Theorem 1. The main difference is
\begin{eqnarray}
n^{*} \leq MG,
\end{eqnarray}
since there are only $MG$ transmit antennas in the network, i.e. the maximum number of transmit symbols in an alignment set is limited to $MG$.
Thus, the maximum value of $n^{*}$ is limited to $\overline{N}=\textrm{min}\{N, MG\}$. For the proof of Theorem 1, the maximum value of $n^{*}$ is also limited to $\textrm{min}\{N, MK\}$, where $MK$ is the number of transmit antennas in $K$-user $M \times 1$ MISO IC.
However, we just assumed that $n^{*}$ is limited to $N$ since
\begin{eqnarray}
MK > M \cdot \underset{\gamma = \left\lfloor\sqrt{\frac{K}{M}}\right\rfloor, \left\lceil\sqrt{\frac{K}{M}}\right\rceil}
{\mathrm{argmin}} M\gamma+\frac{K}{\gamma},
\end{eqnarray}
where the right-hand side means the value of $n^{*}$ that results in the maximum linear sum DoF with sufficiently large $N$.
\end{proof}



If extra preset modes at the receivers more than the number of transmit antennas are not exploited, fully connected interfering broadcast channels only achieve the linear sum DoF as a single cell sum DoF, i.e., a total of $\frac{MK}{M+K-1}$ DoF, not to be affected by inter-cell interference (illustrated by the dotted line in Fig. \ref{IBC}).
Thus, previous work on the sum DoF characterizations for downlink cellular networks (e.g. macro-femto cellular networks \cite{cespedes2}, partially connected cellular networks \cite{cespedes3}-\cite{wei}) assumes that the base stations are able to do data sharing between them to serve the users with all the base stations in their proximity.
In the cellular networks, the sum DoF can be increased by data sharing between the base stations
when the users have enough preset modes to be served by contiguous base stations.
However, Corollary 3 shows that the linear sum DoF for cellular networks can be increased by extra preset modes without the help of data sharing between base stations.
To exploit extra preset modes at the receivers without data sharing, the alignment sets are constructed to include the transmit symbols from the multiple base stations. These symbols are aligned at the other cell's users and have  independent signal subspaces at multiple base stations' users, while a general approach of interference alignment for cellular networks is 
to align transmit symbols from a single base station at all of the other cell's users.
By a similar approach to exploit extra preset modes in Theorem 1 and Corollary 3, partially connected cellular networks could achieve larger sum DoF than fully connected cellular networks to take advantage of their partial connectivity.

We also extend Theorem 1 to fully connected interfering multiple access channel, i.e., uplink cellular networks, by considering the effect of transmitter cooperation for the IC scenario with no CSIT.

\begin{figure}[t]
    \centerline{\includegraphics[width=8.0cm]{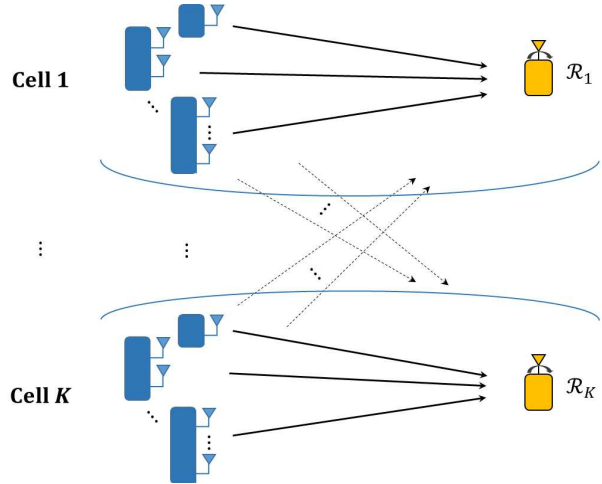}}
    \caption{$K$-cell fully connected interfering multiple access channels, where $M$ transmit antennas exist in each cell.}
    \label{IMA}
    \vspace{-3mm}
\end{figure}

\textbf{Corollary 4}:
For $K$-cell fully connected interfering multiple access channels which are illustrated in Fig. \ref{IMA}, where transmitters in each cell have total $M$ transmit antennas and
each receiver is equipped with a single reconfigurable antenna with $N$ preset modes, the linear sum DoF is also upper-bounded by
(\ref{eq:DoF_n}) as in the $K$-user $M\times1$ MISO IC.

\begin{proof}
Without CSIT, there is no additional DoF gain by the transmitter cooperation, i.e. a single transmitter with $M$ transmit antennas is equivalent to $M$ transmitters with a single transmit antenna in terms of DoF perspective. Thus, fully connected interfering multiple access channels can be seen as $M \times 1$ MISO IC if 
\begin{eqnarray}
\sum\limits_{s \in \mathcal{T}^{k}} M_{s} = M,
\end{eqnarray}
for $k \in [1:K]$, where $\mathcal{T}^{k}$ is the set of transmitters in Cell $k$ and $M_{s}$ denotes the number of transmit antennas for transmitter $s$.
\end{proof}

\section{Conclusion}
\label{sec_con}
This work mainly introduces the theoretical DoF bound for the blind interference technique with reconfigurable antenna switching, where a set of transmit symbols need to be aligned only at their common unintended receivers. In this paper, we derive the linear sum DoF upper-bound for the $K$-user $M \times 1$ MISO IC with reconfigurable antenna at the receivers, and without any channel knowledge at the transmitter, which is induced from the characteristic of the channel matrix with reconfigurable antenna switching. This approach is extended to cellular networks, thus we show that the linear sum DoF can be increased by using sufficient number of preset modes without data sharing between base stations. Moreover, we claim that the linear sum DoF upper-bound is achievable by modifying the existing achievable scheme for MISO IC when certain conditions are satisfied. We also suggest a new achievable scheme for other cases, where certain conditions are not satisfied, but a specific algorithm to make beamforming vectors and preset mode patterns for the achievable scheme are not proposed. A general framework will be investigated as future work.

\begin{appendix}
\label{sec_Lemma3}
\emph{Proof of Lemma 3}:
Suppose that there are $G$ alignment sets which include $n_{g}$ transmit symbols each for $g\in [1:G]$. We can derive the sum DoF upper-bound by calculating the dimension of the desired signal subspace over the dimension of the interfering signal subspace at all the receivers. With $n_g$ transmit symbols, an alignment set occupies $n_g$ dimensions of desired signal subspace, while it also occupies $\big(\left\lceil\frac{n_g}{M}\right\rceil-1\big)n_g$ dimensions at their corresponding receivers. This alignment set is aligned in an 1-dimensional interfering signal subspace at $K-\left\lceil\frac{n_g}{M}\right\rceil$ common unintended receivers. Thus, the dimension of the signal subspace occupied by the alignment set with $n_g$ transmit symbols can be expressed as
\begin{eqnarray}
f(n_g)&=&n_g, \\
h(n_g)&=&\big(\left\lceil\frac{n_g}{M}\right\rceil-1\big)n_g+K-\left\lceil\frac{n_g}{M}\right\rceil,
\end{eqnarray}
where $f(n_{g})$ and $h(n_{g})$ denote the dimensions of the desired signal subspace and interfering signal subspace at all the receivers, respectively, which are occupied by an alignment set with $n_{g}$ transmit symbols. Then, the sum DoF upper-bound with $G$ alignment sets is given by
\begin{eqnarray}
\frac{1}{m} \sum\limits_{j=1}^{K}d_{j}(m) &\leq& \frac{\sum\nolimits_{g=1}^{G}f(n_g) \cdot K}{\sum\nolimits_{g=1}^{G}f(n_g) + \sum\nolimits_{g=1}^{G}h(n_g)} \\
&\stackrel{(a)}{\leq}& \frac{f(n_1) \cdot K}{f(n_1)+h(n_1)} \\
&=& \frac{n_{1}K}{K+\left\lceil\frac{n_{1}}{M}\right\rceil(n_{1}-1)},
\end{eqnarray}
where Step $(a)$ comes from the assumption that $n_1$ is the optimal cardinality of the alignment set as
\begin{eqnarray}
\,\,\,\,\, \frac{f(n_1)}{f(n_1)+h(n_1)} \geq \frac{f(n_g)}{f(n_g)+h(n_g)},
\,\,\,\,\, g \in [2:G].
\end{eqnarray}
It means that the sum DoF upper-bound can be maximized when all alignment sets have the optimal number of elements. This completes the proof. $\blacksquare$
\end{appendix}



\end{document}